\documentclass{article}

\usepackage[margin=4cm]{geometry}
\usepackage{graphicx}
\usepackage[raggedright,tight]{subfigure}  
\usepackage{mathrsfs}

\newcommand{\ket}[1]{\left\vert #1 \right\rangle}
\newcommand{\bra}[1]{\left\langle #1\right\vert}

\begin{document}

\title{\textbf{On the gauge of the natural lineshape}}
\author{\normalsize{Adam Stokes}}

\date{\normalsize{\today}}

\maketitle

\begin{abstract}
We use a general formulation of nonrelativistic quantum electrodynamics in which the gauge freedom is carried by the arbitrary transverse component of the the Green's function for the divergence operator to calculate the natural lineshape of spontaneous emission, thus discerning the full dependence of the result on the choice of gauge. We also use a representation of the Hamiltonian in which the virtual field associated with the atomic ground state is explicitly absent. We consider two processes by which the atom is excited; the first is resonant absorption of incident radiation with a sharp line. This treatment is then adapted to derive a resonance fluorescence rate associated with the Lamb line in atomic hydrogen. Second we consider the atom's excitation due to irradiation with a laser pulse treated semi-classically. An experiment could be used to reveal which of the calculated lineshape distributions is closest to the measured one. This would provide an answer to a question of fundamental importance; how does one best describe atom-radiation interactions with the canonical formalism?
\end{abstract}

\maketitle
 
\section{Introduction}\label{1}

Using conventional theory the natural lineshapes predicted with the minimal coupling and multipolar Hamiltonians are different. This problem was investigated in detail by Power and Zienau, who found that the predictions of the multipolar Hamiltonian were in better agreement with the experiment of Lamb \cite{PandZ,lamb}. It is now well understood that these formally different Hamiltonians correspond to different choices of gauge \cite{woolley1,me,babiker,cohen1}. One must conclude that the dependence of the lineshape on the form of the Hamiltonian is a type of gauge dependence. 

It has been shown that both the minimal coupling and multipolar formulations are particular instances of a more general framework in which the gauge freedom is carried by the arbitrary transverse component of the the Green's function satisfying $\nabla\cdot{\bf g}({\bf x},{\bf x}') = \delta({\bf x}-{\bf x}')$ \cite{woolley1,me}. The Hamiltonian in an arbitrary gauge ${\rm g}$ can be obtained by acting with a unitary gauge fixing transformation $U_{{\rm g}}$ on the Coulomb gauge Hamiltonian $H$ viz. $H_{\rm g} = U_{{\rm g}}HU_{{\rm g}}^{-1}$.

The difference in the lineshapes found using two different gauges is caused by a difference in the physical meanings of the corresponding canonical momenta, which do not commute with $U_{\rm g}$. Ubiquitously in practical calculations throughout QED one starts by partitioning an atom-field Hamiltonian into free and interacting parts based on its superficial appearance as a function of the canonical operators. The gauge dependence of the canonical momenta gives rise to a concurrent gauge dependence of the atom, field and interaction components of the Hamiltonian, as well as the bare atom-photon states and ultimately the lineshape prediction \cite{woolley1}.

This notwithstanding, if one {\em defines} the lineshape observable in terms of the bare states and canonical momenta of a specific gauge ${\rm g}$, the same result can be found in a gauge ${\rm g}'$ by identifying the same physical observables and states as were used in ${\rm g}$. For example, if in the Coulomb gauge we had a state $\ket{\psi}$ and an observable represented by $O$, the same physical state and observable in the gauge-${\rm g}$, are represented by $U_{\rm g}\ket{\psi}$ and $U_{\rm g}OU_{\rm g}^{-1}$ respectively. With this understood, one can see that a definition of the lineshape in terms of some set of {\em physical} observables and states is an implicitly {\em gauge invariant} definition. The task at hand is that of determining the representation in which the canonical operators and bare states happen to represent the physical observables and states that give rise to the correct definition. Such a determination may be possible by comparison of the theoretical predictions with experiment. This in turn would allow one to determine when starting with a canonical partitioning of the Hamiltonian, which representation gives the best description of atom-radiation interactions. Of course, the above reasoning is valid only if there exists with certainty, {\em some} gauge in which the canonical momenta and bare states are the right ones. Failing this one would be forced to seek an entirely different method of calculation altogether \cite{woolley2}.

Previous attempts at resolving the lineshape paradox have been limited to considerations of the Coulomb and Poincar\'e gauges alone. Rather than restrict our attention to two particular cases, the aim in this paper is to investigate the theory of the natural lineshape within the more general ${\rm g}$-gauge formalism and discern its full dependence on the choice of gauge. We will also use a unitary transformation $U_{\rm s}$ similar to those found in Refs. \cite{drummond,me2}, which to first order in the atom-field coupling removes the virtual field associated with the atomic ground state. The resulting representation gives rise to another unique lineshape. 

There are five sections to this paper. In Sec. \ref{2} we introduce the arbitrary ${\rm g}$-gauge Hamiltonian and use it to calculate the various components of the lineshape using the formal theory of radiation damping \cite{heitler,cohen2}. We discuss the necessary and sufficient conditions for the results to be independent of ${\rm g}$. In Sec. \ref{3} we develop a representation in which the virtual field associated with the atomic ground state is explicitly absent. In subsequent sections the lineshape calculations are extended to describe the atom's excitation, it having been noticed some time ago that the lineshape will depend on the details of this procedure \cite{PandZ,low}. In Sec. \ref{4} we describe the preparation of the initially excited atomic state by resonant absorption of incident radiation with a sharp line. With a slight modification the same result can be used to describe the Lamb transition in atomic hydrogen in a way relevant to Lamb's early experiments \cite{PandZ,lamb}. In Sec. \ref{5} we use a simplified treatment whereby the atom is excited by a laser pulse treated semi-classically. We conclude in Sec. \ref{6} with a brief summary of our results.

\section{The natural lineshape in an arbitrary gauge}\label{2}

\subsection{The arbitrary gauge Hamiltonian} 

We consider the case of a single electron with canonical operators ${\bf r}$ and ${\bf p}$ satisfying $[{\bf r},{\bf p}]=i$ bound in an external Coulomb potential and coupled to a transverse radiation field with canonical field operators ${\bf A}_{\rm T}$ and ${\bf \Pi}_{\rm T}$ satisfying $[A_{{\rm T},i}({\bf x}),\Pi_{{\rm T},j}({\bf x}')]=i\delta_{ij}^{\rm T}({\bf x}-{\bf x}')$. In the gauge-${\rm g}$, the longitudinal vector potential is given by \cite{woolley1,me}
\begin{eqnarray}\label{aL}
{\bf A}_{\rm L}= \nabla \chi_{\rm g}({\bf x},[{\bf A}_{\rm T}]),
\end{eqnarray}
where 
\begin{eqnarray}\label{chi}
\chi_{\rm g}({\bf x},[{\bf A}_{\rm T}]) \equiv \int d^3x' \, {\bf g}({\bf x}',{\bf x})\cdot{\bf A}_{\rm T}({\bf x}')
\end{eqnarray}
and ${\bf g}({\bf x},{\bf x}')$ satisfies $\nabla\cdot{\bf g}({\bf x},{\bf x}') = \delta({\bf x}-{\bf x}')$. We define the ${\rm g}$-gauge polarization field as
\begin{eqnarray}\label{Pg}
{\bf P}_{\rm g}({\bf x}) \equiv -\int d^3x'\, {\bf g}({\bf x},{\bf x}')\rho({\bf x}') = e{\bf g}({\bf x},{\bf r})
\end{eqnarray}
where $\rho({\bf x}) = -e\delta({\bf x}-{\bf r})$ is the charge density associated with the electron. In relativistic QED a Coulomb gauge to ${\rm g}$-gauge fixing transformation can be defined by \cite{me}
\begin{eqnarray}\label{ug}
U_{\rm g} \equiv \exp\bigg(i\int d^3x \, \chi_{\rm g}({\bf x},[{\hat {\bf A}}_{\rm T}])\rho({\bf x}) \bigg) = \exp\bigg(-i\int d^3x \, {\bf P}_{\rm g}({\bf x})\cdot {\bf A}_{\rm T}({\bf x}) \bigg).
\end{eqnarray}
$U_{\rm g}$ has the form of the Power-Zienau-Woolley (PZW) transformation, but note that the polarization field in Eq. (\ref{ug}) is essentially arbitrary and need not be identified with the usual multipolar polarization field. In the non-relativistic setting a general ${\rm g}$-gauge Hamiltonian can be defined by $H_{\rm g} \equiv U_{\rm g}HU_{\rm g}^{-1}$ where $H$ is the Coulomb gauge Hamiltonian and ${\bf g}({\bf x},{\bf x}')$ is chosen so that $[[[H_0,\chi_{\rm g}],\chi_{\rm g}],\chi_{\rm g}] = [[V_1,\chi_{\rm g}],\chi_{\rm g}]= [V_2,\chi_{\rm g}]=0$, where $V_i$ denotes the interaction component of $H$ up to order $i=1,2$ in the coupling ($e$) \cite{woolley3}. The ${\rm g}$-gauge Hamiltonian $H_{\rm g}$ then has the following interaction component \cite{woolley1}
\begin{eqnarray}\label{Vg}
V_{\rm g} = {e\over 2m}\bigg[{\bf p}\cdot{\bf A}({\bf r})+ {\bf A}({\bf r})\cdot{\bf p} + e{\bf A}({\bf r})^2\bigg] + \int d^3x \,  \bigg[{\bf \Pi}_{\rm T}({\bf x})\cdot {\bf P}_{\rm g}({\bf x}) + {1\over 2}{\bf P}_{\rm g}({\bf x})^2 \bigg],
\end{eqnarray}
where ${\bf A}\equiv{\bf A}_{\rm T}+{\bf A}_{\rm L}$.

\subsection{Calculation of the lineshape}

The natural lineshape is the frequency distribution of radiation spontaneously emitted by an atom in an excited state. Assuming the atom is in the first excited state $\ket{e;0}$ at $t=0$, where $\ket{0}$ denotes the photon vacuum, we calculate the long-time squared amplitude $\vert b_{g;{\bf k}\lambda}(\infty)\vert^2$, representing the probability of finding the atom in the ground state $\ket{g}$ and a photon $\ket{{\bf k}\lambda}$ with frequency $\omega_k$ present upon measurement. The lineshape is defined as
\begin{eqnarray}\label{S}
S(\omega_k) \equiv \rho(\omega_k) \left({L \over 2\pi}\right)^3 \int d\Theta \sum_\lambda \vert b_{g;{\bf k}\lambda}(\infty)\vert^2,
\end{eqnarray}
where $\rho(\omega_k) =\omega_k^2$ is the density of field modes, the sum is over polarisations belonging to a given direction, and the integration is over all directions. To derive an expression for $\vert b_{g;{\bf k}\lambda}(\infty)\vert^2$ we start with the traditional method \cite{dirac} of considering the variation of the coefficients $b_f(t)\equiv\langle f\vert\psi (t)\rangle$ associated with a Hamiltonian $H=H_0+V$, in the interaction picture. Following \cite{heitler} we introduce the Fourier transform
\begin{eqnarray}\label{G}
b_f(t)=-{1\over 2\pi i}\int d\omega \, G_{fi}(\omega)e^{i(\omega_n-\omega)t},
\end{eqnarray}
where $G(\omega)=1/(\omega-H)$ is the resolvent operator, and $\ket{i}$ is the initial state at $t=0$. Next we introduce the shift operator $U(\omega)$ satisfying
\begin{eqnarray}\label{U}
U(\omega) = V+VG_0(\omega)U(\omega),
\end{eqnarray}
where $G_0(\omega)=1/(\omega-H_0)$ is the free resolvent \cite{cohen2,roman}. Formal manipulations lead in a straightforward fashion to the result \cite{heitler}
\begin{eqnarray}\label{prob}
\vert b_f(\infty)\vert^2 = {\vert U_{fi}(\omega_f)\vert^2 \over (\omega_f-\omega_i-\Delta\omega(\omega_f))^2+(\Gamma(\omega_f)/2)^2},
\end{eqnarray}
where \cite{heitler,cohen2}
\begin{eqnarray}\label{gamdeltaom}
\Gamma(\omega) = 2\pi\sum_n \vert U_{ni}(\omega)\vert^2\delta(\omega-\omega_n), ~~~~~ \Delta\omega(\omega) = {\mathscr P} \sum_n {\vert U_{ni}(\omega_n)\vert^2 \over \omega-\omega_n},
\end{eqnarray}
with ${\mathscr P}$ denoting the principal value. In the case $\ket{f}=\ket{g;{\bf k}\lambda}$ and $\ket{i} = \ket{e;0}$, Eq. (\ref{prob}) defines the amplitude in Eq. (\ref{S}). In practice an explicit solution can only be found using an expansion of $U(\omega)$ in powers of $V$. Therefore, in order to go further we make the first approximation $U_{nm}(\omega)\approx V_{nm}$. The components of the lineshape now only depend on matrix elements of the form $\bra{n;{\bf k}\lambda}V\ket{e;0}$. Using Eq. (\ref{Vg}) we find
\begin{eqnarray}\label{element}
\hspace*{-1.5cm}\bra{n;{\bf k}\lambda}V_{\rm g}\ket{e;0} &=&  {e g_k \over m} {\bf e}_{{\bf k}\lambda} \cdot  [{\bf p}e^{-i{\bf k}\cdot{\bf r}}]_{ne} ~+~ ie\omega_k g_k \, \, {\bf e}_{{\bf k}\lambda} \cdot {\tilde {\bf g}}_{ne}({\bf k},{\bf r}) \nonumber \\ && + {eg_k\over 2m} \bigg[ {\bf p}\cdot \nabla_r \big\{{\bf e}_{{\bf k}\lambda} \cdot {\tilde {\bf g}}({\bf k},{\bf r})\big\} + \nabla_r \big\{{\bf e}_{{\bf k}\lambda} \cdot {\tilde {\bf g}}({\bf k},{\bf r})\big\}\cdot {\bf p}\bigg]_{ne},
\end{eqnarray}
where
\begin{eqnarray}\label{gtilde}
g_k \equiv  \sqrt{1 \over  2\omega_k L^3 }, ~~~~~ {\tilde {\bf g}}({\bf k},{\bf r}) \equiv \int d^3x \, {\bf g}({\bf x},{\bf r})e^{-i{\bf k}\cdot{\bf x}}.
\end{eqnarray}
The matrix element in Eq. (\ref{element}) is clearly explicitly dependent on the gauge choice ${\rm g}$. Choosing the Coulomb gauge and making the electric dipole approximation (EDA) $e^{i{\bf k}\cdot{\bf r}}\approx 1$, we obtain the following approximate expressions for respectively, the numerator, level shift and decay rate in Eq. (\ref{prob})
\begin{eqnarray}\label{cg}
&& \vert U_{e;0,g;{\bf k}\lambda} (\omega_g + \omega_k) \vert^2 \approx  {e^2\over m^2} {1\over 2\omega_k L^3} \vert {\bf p}_{eg} \cdot {\bf e}_{ {\bf k}\lambda} \vert^2, \nonumber \\
&& \Delta \omega(\omega_g+\omega_k) \approx  {\mathscr P}  \sum_n \sum_{{\bf k}'\lambda'} {e^2 \over m^2} {1\over 2\omega_{k'} L^3} {\vert {\bf p}_{ne}\cdot{\bf e}_{{\bf k}'\lambda'}\vert^2 \over \omega_k -\omega_{ng}-\omega_{k'} }, \ \nonumber \\
&& \Gamma(\omega_g+\omega_k) \approx 2\pi \sum_n \sum_{{\bf k}'\lambda'} {e^2 \over m^2} {1\over 2\omega_{k'} L^3}\vert{\bf p}_{ne}\cdot{\bf e}_{{\bf k}'\lambda'}\vert^2\delta(\omega_k -\omega_{ng} -\omega_{k'}), 
\end{eqnarray}
where $\omega_{ng}\equiv\omega_n-\omega_g$. Similarly using the Poincar\'e gauge and EDA one obtains
\begin{eqnarray}\label{pg}
&& \vert U_{e;0,g;{\bf k}\lambda} (\omega_g + \omega_k) \vert^2 \approx   {\omega_k\over 2L^3} \vert {\bf d}_{eg} \cdot {\bf e}_{ {\bf k}\lambda}  \vert^2, \nonumber \\
&& \Delta \omega(\omega_g+\omega_k) \approx {\mathscr P}  \sum_n \sum_{{\bf k}'\lambda'}  {\omega_{k'}\over 2 L^3} {\vert {\bf d}_{ne}\cdot{\bf e}_{{\bf k}'\lambda'}\vert^2 \over \omega_k -\omega_{ng}-\omega_{k'} }, \ \nonumber \\
&& \Gamma(\omega_g+\omega_k) \approx 2\pi \sum_n \sum_{{\bf k}'\lambda'}  {\omega_{k'} \over 2 L^3}\vert{\bf d}_{ne}\cdot{\bf e}_{{\bf k}'\lambda'}\vert^2\delta(\omega_k -\omega_{ng} -\omega_{k'}), 
\end{eqnarray}
where ${\bf d} = -e{\bf r}$. We see that, as is implied by Eq. (\ref{element}), the expressions in (\ref{cg}) are not the same as those in (\ref{pg}). 

\subsection{The approximations ensuring gauge invariance}

We review here the various approximations, which can be used to reproduce previous results and eliminate the dependence of the lineshape on the choice of gauge. We note first that three approximations have already been used; first the limiting value $t\to\infty$ ensures the level shift and decay rate in Eq. (\ref{prob}) are evaluated at $\omega_f  = \omega_g+\omega_k$. Second an approximation $U_{nm}(\omega) \approx V_{nm}$ is used for the matrix elements of the shift operator. Finally the EDA is used to give the eventual expressions in (\ref{cg}) and (\ref{pg}). In the discussion that follows it is assumed that all three of these approximations have been made. 

Crucial in ensuring gauge invariance is the further approximation of insisting that the emission process conserves energy i.e. that $\omega_k = \omega_{eg}$. In the decay rate $\Gamma$ the delta function then ensures that the matrix element of $V_g$ is evaluated on-energy-shell. It is a standard result that such matrix elements are quite generally invariant for two Hamiltonians related by a unitary transformation of the form $U=\exp(ieS)$ \cite{craig,woolley2}. The invariance of $\Gamma$ can easily be verified explicitly for the Coulomb and Poincar\'e gauges by using the relation 
\begin{eqnarray}\label{posmom}
{\bf p}_{nm} = im\omega_{nm}{\bf r}_{nm}
\end{eqnarray}
between matrix elements of position and momentum in the free energy basis. The result is nothing but the well-known first order decay rate found using Fermi's golden rule \cite{cohen2,craig}.

Turning our attention to the level shift $\Delta \omega$, it is easy to check that the imposition of energy conservation alone does not suffice to ensure gauge invariance. For this, one must also add to $\Delta \omega$ the contribution $\bra{e;0}V_{\rm g}\ket{e;0}$, which produces the total shift
\begin{eqnarray}\label{totshift}
\Delta \omega_{\rm total} &=& \bra{e;0}V_{\rm g}\ket{e;0} + \sum_n {\vert\bra{n}V_{\rm g}\ket{e;0}\vert^2 \over \omega_e-\omega_n} .
\end{eqnarray}
This is the same on-energy-shell shift in energy of the excited state $\ket{e;0}$ as is obtained through second order stationary perturbation theory. Like $\Gamma$ it is invariant for two Hamiltonians related by a unitary transformation $U=\exp(ieS)$ \cite{craig}. As with $\Gamma$ this invariance is easily verified for the Coulomb and Poincar\'e gauges. In the Coulomb gauge the additional contribution in Eq. (\ref{totshift}) comes from the $e^2{\bf A}^2/2m$ part of the interaction Hamiltonian to give a total shift
\begin{eqnarray}\label{totshiftcg}
\Delta \omega_{\rm total} &=& \sum_{{\bf k}\lambda} {e^2 \over m} {1\over 2\omega_k L^3} \left( {\vert  {\bf e}_{{\bf k}\lambda}\vert^2 \over 2} - \sum_n  {1\over m} {\vert {\bf p}_{ne}\cdot{\bf e}_{{\bf k}\lambda} \vert^2 \over \omega_{ne}+\omega_k }\right).
\end{eqnarray}
In the Poincar\'e gauge the additional contribution comes from the polarisation field term $e^2\vert {\bf r}\delta^{\rm T}\vert^2/2$ corresponding to the last term in Eq. (\ref{Vg}). Hence, the total shift is
\begin{eqnarray}\label{totshiftpg}
\Delta \omega_{\rm total} &=& \sum_n \sum_{{\bf k}\lambda} {1\over 2 L^3} \vert{\bf d}_{ne} \cdot {\bf e}_{{\bf k}\lambda}\vert^2 \left( 1- {\omega_k \over \omega_{ne}+\omega_k}\right).
\end{eqnarray}
Using the expansion
\begin{eqnarray}\label{expansion}
{1\over \omega_{ne}  +\omega_k} = {1\over \omega_k} + {\omega_{ne} \over \omega_k^2} +{\omega_{ne}^2 \over \omega_k^3} - {\omega_{ne}^3 \over \omega_k^3(\omega_{ne}+\omega_k)}
\end{eqnarray}
and Eq. (\ref{posmom}) it is straightforward to show that (\ref{totshiftcg}) and (\ref{totshiftpg}) are identical \cite{craig}. Neglecting in $\Delta \omega_{\rm total}$, all contributions not dependent on the state of the electron, and removing the electron self-energy contribution through mass renormalization gives the standard non-relativistic Lamb shift \cite{craig}
\begin{eqnarray}\label{LS}
\Delta \omega_{\rm LS} = \sum_n \sum_{{\bf k}\lambda} {e^2 \over m^2} {1\over 2\omega_k^2 L^3} {\omega_{ne} \vert{\bf p}_{ne}\cdot{\bf e}_{{\bf k}\lambda}\vert^2 \over \omega_{ne}+\omega_k }.
\end{eqnarray}

\subsection{Discussion}

The energy conservation condition $\omega_k = \omega_{eg}$ has been justified on the grounds that $\Gamma(\omega)$ and $\Delta\omega(\omega)$ do not vary appreciably over the interval $w\gg\Gamma,\Delta\omega$ centered at $\omega_e$, and they can therefore be evaluated at $\omega_e$ to within sufficient accuracy \cite{cohen2}. However, this argument is somewhat ruined by the gauge arbitrariness of the matrix elements of $U(\omega)$, more precisely, the required slow variations of $\Gamma(\omega)$ and $\Delta\omega(\omega)$ cannot be guaranteed irrespective of ${\bf g}$. At the same time it is clear that without energy conservation, and the ad hoc modification of the level shift given in Eq. (\ref{totshift}), the denominator in Eq. (\ref{prob}) is gauge dependent.

Nevertheless, for most ``sensible" choices of gauge energy conservation as an approximation may certainly be valid and good. In such cases its use would be of little or no practical significance. Moreover, the modification of adding to $\Delta \omega$ the contribution $\bra{e;0}V_{\rm g}\ket{e;0}$ coming from the first order in perturbation theory gives a clear prescription by which the level shift can be given an unambiguous interpretation as the Lamb shift of the excited atomic state.

The numerator $\vert U_{fi}(\omega_f)\vert^2$ in Eq. (\ref{prob}) is more troublesome. It is typically neglected altogether, or otherwise evaluated on-energy-shell, yielding the gauge invariant result $\Gamma/2\pi$. Either procedure can only be justified on the grounds that the numerator's dependence on $\omega_k$ is sufficiently slow so as to be undetectable when compared with the denominator. If this is not the case then it should be possible to determine with an experiment, which form of the Hamiltonian produces the most accurate lineshape prediction.

\section{The lineshape after removal of the virtual field}\label{3}

\subsection{The Hamiltonian and lineshape in the symmetric representation}

The gauge dependence of the lineshape stems from the gauge dependent specification of bare states between which transitions are supposed to occur. In general the free energy $H_0$ does not commute with the interaction $V$ and so it is not conserved. It may be argued that the notion of free energy is useful only insofar as it is conserved. From a more physical perspective, the common interpretation of the atom is one in which it is surrounded by a virtual cloud of photons that are continually emitted and reabsorbed. These virtual processes give rise to shifts in the energy of the bare atomic states. In describing spontaneous emission it may be preferable that the virtual cloud be included implicitly in the bare states. For the ground state such a description is possible via diagonalization of the Hamiltonian, to the extent that the bare atom-photon state $\ket{g;0}$ coincides with the true ground state. In such a representation the free energy becomes a symmetry of the Hamiltonian $[H_0,H]=0$, and consequently is perhaps a more legitimate source of bare states for the lineshape calculation.

\begin{center}
\begin{figure*}[t]\label{fig1} 
\subfigure[\hspace*{-0mm}]
{\includegraphics[width=0.48\textwidth]{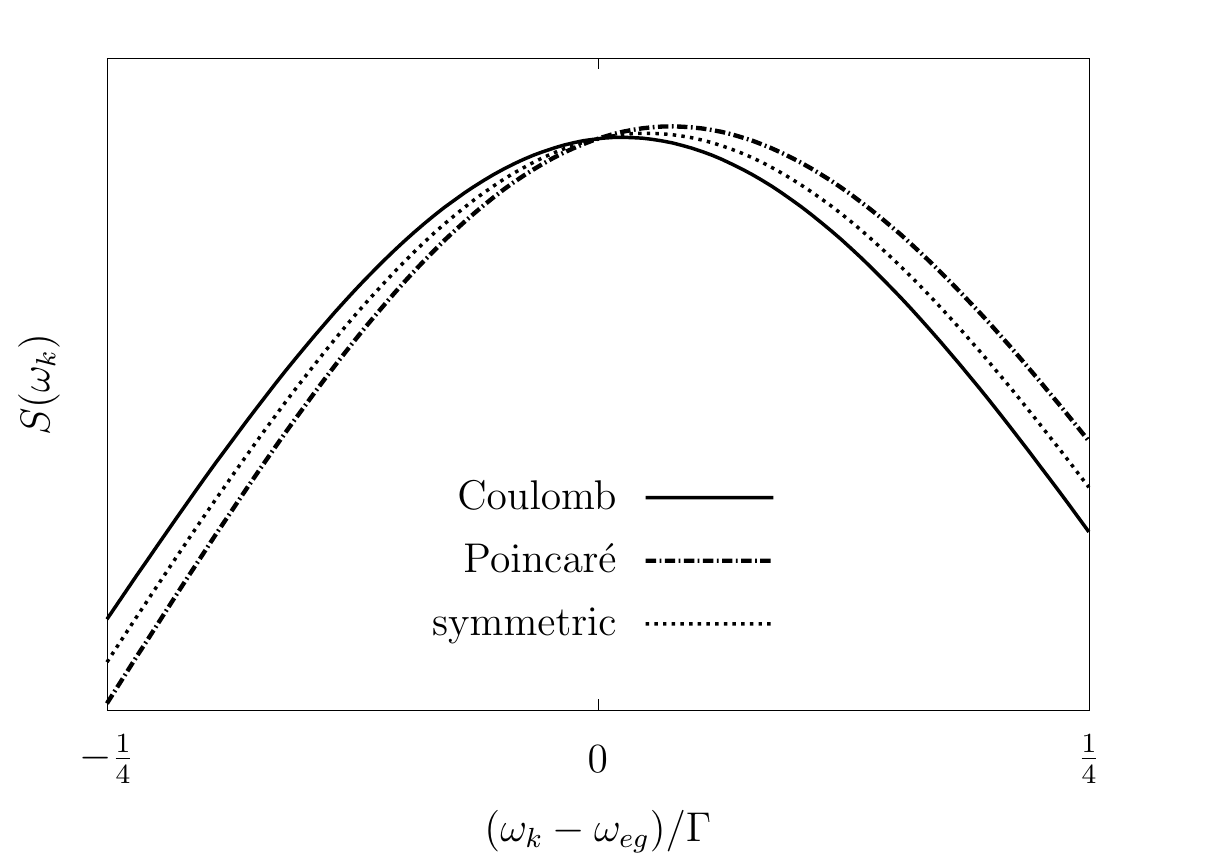}\label{f1}}
\subfigure[\hspace*{-3mm}]
{\includegraphics[width=0.48\textwidth]{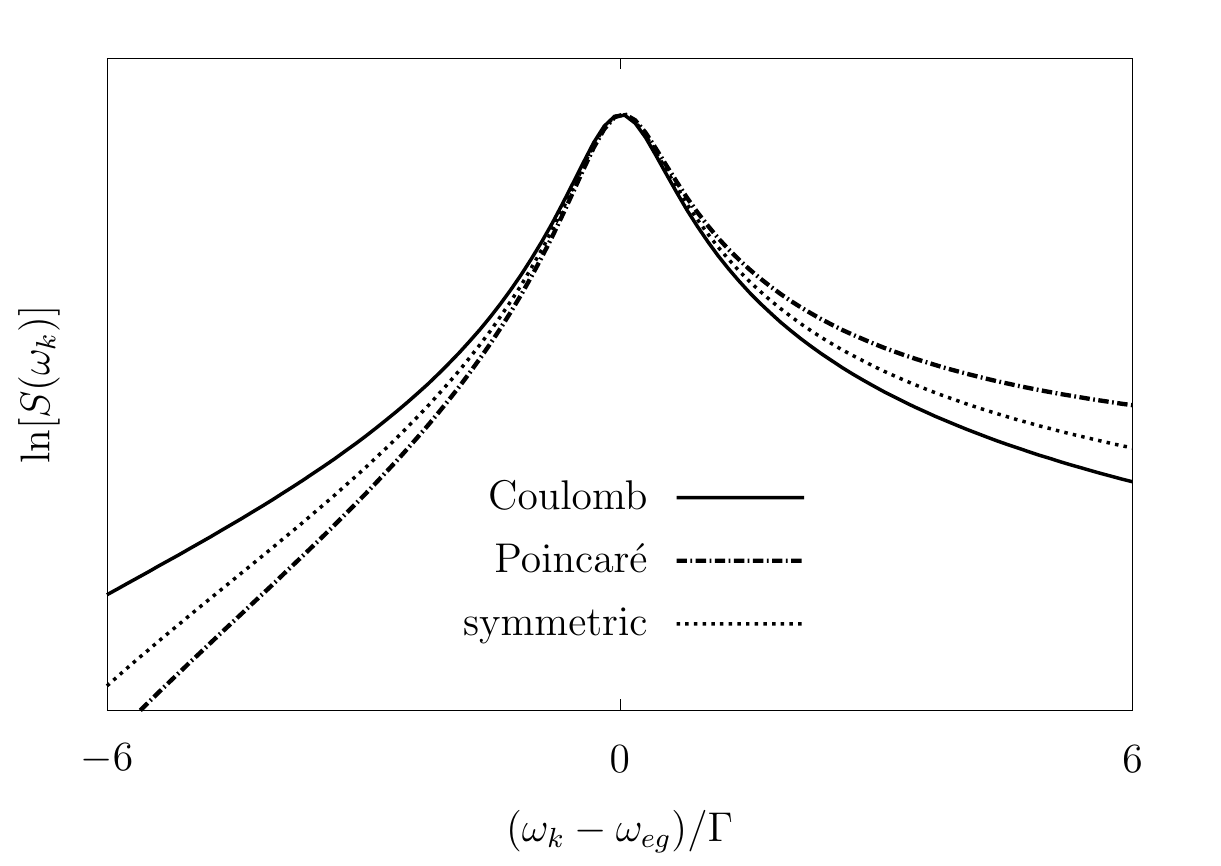}\label{f2}}
\vspace*{-0.2cm}
\caption{\small{The lineshapes associated with the Coulomb gauge, Poincar\'e  gauge and the symmetric representation with $\Gamma = \omega_{eg}/10$. In each case the Lamb shift $\Delta \omega_{\rm LS}$ has been suppressed. In \ref{f1} $S(\omega_k)$ is plotted, whereas in \ref{f2} $\ln[S(\omega_k)]$ is plotted. Since the coupling in the symmetric representation is a symmetric mixture of the Coulomb gauge and Poincar\'e gauge couplings, the corresponding curve interpolates between the curves associated with these gauges.}}
\end{figure*}
\end{center}

Starting in the Coulomb gauge and EDA we define the unitary operator
\begin{eqnarray}\label{Uv}
U_{\rm s} \equiv \exp\bigg[ -\sum_{{\bf k}\lambda} \sum_{nm} ig_k ({\bf e}_{{\bf k}\lambda}\cdot{\bf d}_{nm}) \alpha_{k,nm} \ket{n}\bra{m} (a_{{\bf k}\lambda}^\dagger + a_{{\bf k}\lambda})\bigg],
\end{eqnarray}
where the $\ket{n}$ are Coulomb gauge bare atomic states, $a_{{\bf k}\lambda}$ is the Coulomb gauge photon annihilation operator for the mode ${\bf k}\lambda$, ${\bf e}_{{\bf k}\lambda}$ is the corresponding unit polarization vector, and $g_k$ is defined in Eq. (\ref{gtilde}).This transformation is the extension to the case of a general multi-level atom, of those found in Refs. \cite{drummond,me2}. It is different to the one found in \cite{zubiary}. The term $\alpha_{k,nm}$ is chosen so as to eliminate the energy non-conserving terms in the linear part of the Coulomb gauge interaction Hamiltonian. The appropriate choice being
\begin{eqnarray}\label{alpha}
\alpha_{k,nm} = {\vert\omega_{nm}\vert \over \omega_k + \vert \omega_{nm} \vert}.
\end{eqnarray}
It is important to note that $U_{\rm s}$ only eliminates the energy non-conserving terms to first order in the coupling, and within the EDA. The resultant Hamiltonian $H_{\rm s}\equiv U_{\rm s}HU_{\rm s}^{-1}$ has to first order in $e$, the interaction component
\begin{eqnarray}\label{Vv}
V_{\rm s} = \sum_{{\bf k}\lambda } \sum_{n,m}^{n>m}  g_{{\bf k} \lambda,nm} \,  {2 (\omega_{nm} \omega_k)^{1/2} \over \omega_{nm} + \omega_k} \, \ket{n}\bra{m} a_{{\bf k}\lambda } + {\rm H.c.}
\end{eqnarray}
where 
\begin{eqnarray}
g_{{\bf k} \lambda,nm} \equiv  -i\left({\omega_{nm} \over 2 L^3}\right)^{1/2} \, ({\bf  e}_{{\bf k}\lambda }\cdot{\bf d}_{nm}).
\end{eqnarray}
The resultant representation symmetrically mixes the Coulomb gauge and Poincar\'e gaue couplings, so we refer to it as the symmetric representation. Terms of $O(e^2)$ have been omitted, because they give rise to $O(e^4)$ contributions in the lineshape; their only contribution is to the term $\bra{e;0}V\ket{e;0}$ in Eq. (\ref{totshift}). Yet we know this level shift is invariant under a unitary transformation $U=\exp(ieS)$, so the total level shift obtained in the new representation must be the same as for the Coulomb gauge. The same is true of the decay rate $\Gamma$, and therefore of the denominator in Eq. (\ref{prob}). The lineshape in the symmetric representation is
\begin{eqnarray}
S(\omega_k) = {{ 4\omega_k^3 \over \omega_{eg}(\omega_{eg} + \omega_k)^2} {\Gamma/2\pi \over (\omega_k-\omega_{eg}-\Delta \omega_{\rm LS})^2 + \Gamma^2/4}}, 
\end{eqnarray}
which is plotted in Figs. \ref{f1} and \ref{f2} along with the Coulomb and Poincar\'e gauge results. The difference in lineshapes between gauges is a result of the differing $\vert U_{fi}(\omega_f)\vert^2$ terms of Eq. (\ref{prob}), which are collected in table \ref{table} for the three main cases.
\begin{table}[t]
\begin{center}
\begin{tabular}{ccccc}  
        \hline \\[-0.2cm]
	 ~~Representation~~  &   ~~ $\vert U_{fi}(\omega_f)\vert^2$ term in second order ($\times 2\pi/\Gamma$)  & 
	\\[0.2cm] \hline \\[-0.2cm]
	Coulomb & ~~$ {\omega_k \over \omega_{eg}} $ &   
	\\ \\
	Poincar\'e & ~~$\left( {\omega_k \over \omega_{eg}} \right)^3$ &   
	\\ \\
	symmetric & ~~${4\omega_k^3 \over \omega_{eg} (\omega_{eg} + \omega_k)^2}$ & 
	\\[0.4cm] \hline 
\end{tabular}
\end{center}
\caption{\small{The frequency dependence of the lineshape numerator $\vert U_{fi}(\omega_f)\vert^2$ of Eq. (\ref{prob}) in different representations.}}\label{table}
\end{table}
%
%
%
%

\newpage

\section{The atom's excitation through resonant absorption}\label{4}

\subsection{Absorption of incident radiation with a sharp line}\label{sharp}

We consider the situation whereby the atom starts in its ground state in the presence of incident radiation with intensity distribution $S$. A primary photon with frequency $\omega_0$ is absorbed by the atom out of $S$ and a photon with frequency $\omega_k$ near $\omega_0$ is emitted. The quantity of interest is the total rate $\gamma$ at which the system leaves the initial state $\ket{i}$. This is obtained using the theory of radiation damping used in the preceding sections. Rather than give a detailed derivation, which can be found in, for example, \cite{heitler}, we will simply state the main results. The rate $\gamma$ is given by
\begin{eqnarray}\label{rate}
\gamma = \sum_n {\vert V_{ni}\vert^2 \over (\omega-\omega_n)^2+\Gamma(\omega_n)^2/4},
\end{eqnarray}
where $\Gamma(\omega)$ is defined in Eq. (\ref{gamdeltaom}) with $U_{nm}(\omega)$ replaced by $V_{nm}$. If the incident radiation is sharp i.e. different from zero only at $\omega=\omega_0$, and the sum in Eq. (\ref{rate}) is only extended over this line we obtain the total rate of resonance fluorescence \cite{heitler}
\begin{eqnarray}\label{rate2}
{\gamma \over L^3}= {S\,\Gamma(\omega_0)\over \omega_0} {\vert \bra{g;{\bf k}_0\lambda}V\ket{e;0} \vert^2 \over (\omega_0 - \omega_{eg})^2 + \Gamma^2 / 4},
\end{eqnarray}
where $\Gamma$ in the denominator is the gauge invariant on-energy-shell decay rate of the excited state. The transition rate is clearly dependent on the form of the interaction and therefore gauge dependent. The rate can be written
\begin{eqnarray}\label{rate3}
\gamma = {S\,\Gamma \vert{\bf e}\cdot{\bf d}_{ge}\vert^2 \over 2} {n(\omega_0,\omega_{eg}) \over (\omega_0-\omega_{eg})^2+\Gamma^2/4},
\end{eqnarray}
where the incident signal, intensity $S$ is polarised along ${\bf e}$. The ``numerator" $n$ differs between the different representations, analogously to the difference in the $\vert U_{fi}(\omega_f)\vert^2$ term of the lineshape in Secs. \ref{2} and \ref{3}. This is summarised in table \ref{table2} for the three main cases.

\begin{table}[h]
\begin{center}
\begin{tabular}{ccccc}  
        \hline \\[-0.2cm]
	 ~~Representation~~  &   ~~$n(\omega_0,\omega_{eg})$  & 
	\\[0.2cm] \hline \\[-0.2cm]
	Coulomb & ~~${\omega_{eg} \over \omega_0}$ &   
	\\ \\
	Poincar\'e & ~~$\left({\omega_0 \over \omega_{eg}}\right)^3$ &   
	\\ \\
	symmetric & ~~${16\omega_{eg}\omega_0^3 \over (\omega_{eg} + \omega_0)^4}$ & 
	\\[0.4cm] \hline 
\end{tabular}
\end{center}
\caption{\small{The frequency dependence of the different resonance fluorescence rates.}}\label{table2}
\end{table}
\vspace*{-0.4cm}

\subsection{The Lamb line in hydrogen}\label{lambline}

The $2s\to1s$ transition in atomic hydrogen has received a great deal of attention over the years \cite{PandZ,lamb,cohen1,fried,bassani}. Transition matrix elements on-energy-shell are gauge invariant, the matrix element of the two photon resonant transition $2s\to1s$ being a particular example. This has been amply verified with semi-classical treatments when complete sets of intermediate states are used \cite{cohen1,fried,bassani}. On the other hand lineshape formulas including radiation damping will not in general produce the same results in each gauge \cite{fried,davidovich}. Either one concludes that the theory of radiation damping is unsatisfactory for a fundamental account of the frequency dependence of spectral lines \cite{woolley2}, or that the question of the correct lineshape is a matter to be resolved empirically \cite{power3}.

The quantity relevant to the experiments of Lamb is the fluorescence rate $\gamma$ out of the metastable state $2s$ in a process of stimulated decay, due to the presence of a microwave signal with frequency $\omega_0$ near the Lamb separation $\omega$ of the $2s\to 2p$ transition. Energy conservation implies that a photon with frequency $\omega_k = \omega+\omega'-\omega_0$ is emitted in the cascade $2s\to 2p\to 1s$, where $\omega'$ is the frequency of the $2p\to 1s$ transition. The spontaneous single photon decay process $2s\to 2p$ is negligible, as is the spontaneous two photon decay process $2s\to 2p\to 1s$. Moreover, the sharp width condition on the $2s$ level in the presence of an incident microwave signal is satisfied for all signal levels, which means the treatment of the preceding section should be valid. The rate $\gamma$ is given by Eq. (\ref{rate2}) with some changes \cite{PandZ}
\begin{eqnarray}\label{rate4}
{\gamma \over L^3}= {S\,\Gamma_{2p,1s}(\omega+\omega'-\omega_0) \over \omega_0}{\vert\bra{2s;{\bf k}_0\lambda}V\ket{2p;0}\vert^2 \over (\omega_0 - \omega)^2 + \Gamma_{2p,1s}^2 / 4}.
\end{eqnarray}
With this we obtain
\begin{eqnarray}\label{rate5}
\gamma = {S\,\Gamma_{2p,1s} \vert{\bf e}\cdot{\bf d}_{2s,2p}\vert^2 \over 2} {n'(\omega_0,\omega,\omega') \over (\omega_0-\omega)^2+\Gamma_{2p,1s}^2/4},
\end{eqnarray}
where as before $n'$ differs between the different representations, as is summarised in table \ref{table3}.

\begin{table}[h]
\begin{center}
\begin{tabular}{ccccc}  
        \hline \\[-0.2cm]
	 ~~Representation~~  &   ~~$n'(\omega_0,\omega,\omega')$  & 
	\\[0.2cm] \hline \\[-0.2cm]
	Coulomb & ~~${\omega + \omega' - \omega_0 \over \omega'}{\omega^2 \over \omega_0^2}$ &   
	\\ \\
	Poincar\'e & ~~$\left({\omega + \omega' - \omega_0 \over \omega'}\right)^3$ &   
	\\ \\
	symmetric & ~~ ${4(\omega+\omega'-\omega_0)^3 \over \omega'(\omega+2\omega'-\omega_0)^2} {4\omega^2 \over (\omega +\omega_0)^2}$ &
	\\[0.4cm] \hline 
\end{tabular}
\end{center}
\caption{\small{The frequency dependence of the Lamb line within different representations.}}\label{table3}
\end{table}

Lamb's experiments yielded a distribution of frequencies $\omega$ as a function of intensity. The relative differences in the distributions $\gamma(\omega)$ associated with the different $n'$ in table \ref{table3} are essentially the same as the relative differences in the lineshapes plotted in Figs. \ref{f1} and \ref{f2}. The Poincar\'e gauge result is indistinguishable from the bare Lorentzian whereas the other curves exhibit small deviations.

\subsection{Discussion}

A good deal of work has been put into clarifying the conditions under which predictions pertaining to the two-photon Lamb transition $2s\to 1s$ in atomic hydrogen are gauge invariant. Within the $S$-matrix formalism the on-energy-shell transition probability amplitude associated with some transition $\ket{i}\to \ket{f}$ is gauge invariant provided complete sets of any intermediate states are retained. The crucial condition in ensuring gauge invariance is energy conservation expressed through the delta function in the equation
\begin{eqnarray}\label{Sm}
S_{fi} &=& \delta_{fi} +\delta(\omega_f-\omega_i)T_{fi},
\end{eqnarray}
in which the $T$-matrix is essentially the level shift operator $U(\omega)$ in Eq. (\ref{U}), evaluated at $\omega=\omega_i$. The element $T_{fi}$ in Eq. (\ref{Sm}) satisfies
\begin{eqnarray}
T_{fi} = V_{fi} + \sum_n {V_{fn}T_{ni} \over \omega_i - \omega_n +i\epsilon},
\end{eqnarray}
with the usual limit $\epsilon\to 0^+$ understood. Woolley has given a proof of the gauge invariance of the on-energy-shell $T$-matrix elements, within the $g$-gauge formalism \cite{woolley3}. Within the radiation damping theory used in Secs. \ref{2}, \ref{3} and \ref{4}, one specifies the initial state at $t_i = 0$. In contrast, the $S$-matrix element $S_{fi}$ gives the amplitude associated with the $\ket{i}\to \ket{f}$ transition in the limit $t_i\to -\infty$ and $t_f\to \infty$. It is only in this {\em doubly} infinite limit that the energy conservation condition crucial in ensuring gauge invariance appears.

For two-photon scattering processes the $S$-matrix in second order yields the Kramers-Heisenberg dispersion relation \cite{craig_p1}. However, a treatment limited to second order is insufficient to describe the exponential decay of the excited atomic state, which gives rise to the decay rate in the denominator of the lineshape. Thus, in order to obtain a lineshape formula from the Kramers-Heisenberg formula one must first add a decay term into the denominator of the resonant contribution \cite{sakuri_p}. Ignoring the non-resonant contribution and restricting one's attention to the apparently dominant $2p$ intermediate state, the Kramers-Heisenberg formula can be used to obtain the results of Sec. \ref{sharp} \cite{PandZ}. Of course in carrying out these steps one breaks the gauge invariance of the matrix element. 

For the two-photon Lamb transition $2s\to 1s$ with emissions ${\bf k}\lambda$ and ${\bf k}'\lambda'$, the second order $S$-matrix element in the Poincar\'{e} gauge and the electric dipole approximation is \cite{craig_p2}
\begin{eqnarray}\label{KH}
S_{fi}={\sqrt{\omega_k\omega_{k'}}\over 2V} {\rm e}_{{\bf k}\lambda}^i {\rm e}^j_{{\bf k}'\lambda'} \sum_n \left({ d^i_{1s,n} d^j_{n,2s}  \over \omega_{n,1s} - \omega_k }+{ d^j_{1s,n} d^i_{n,2s} \over \omega_{n,1s} - \omega_{k'} }\right) \delta(\omega_{2s,1s} -\omega_k - \omega_k')  
\end{eqnarray}
where the repeated spatial indices $i,j$ are assumed to be summed. The matrix element in Eq. (\ref{KH}) is well-known to be the same in the Coulomb and Poincar\'{e} gauges when the full set of intermediate states is retained \cite{cohen1,craig,fried,bassani}. This of course, is just a particular case of the general gauge invariance of the $S$-matrix. In actual fact by appropriately summing all terms in the $S$-matrix one can obtain a lineshape formula for the Lamb line in hydrogen directly, but this still requires the restriction to the single intermediate state $2p$ \cite{powerapp}. One can then derive the results of Sec \ref{lambline}.

Under the condition that predictions must be {\em manifestly} gauge invariant and conserve energy, one is restricted to the $S$-matrix formalism, which is not dynamical. The desire for a dynamical theory of matter-radiation interactions was expressed early on in the works of Heitler \cite{heitler}, and its attractiveness seems to have persisted \cite{bach}. In the radiation damping formalism, considering finite times is equivalent to considering arbitrary energy values $\omega$ in Eq. (\ref{U}). In particular, the well-known quadratic dependence of the lineshape at short-times offers a correction to the bare Lorenztian result \cite{cohen2}.

However, as soon as finite times are considered, predictions pertaining to canonical degrees of freedom will yield different results in different gauges. It is important to recognise that this type of gauge dependence is not merely an artifact of some inevitable perturbative ansatz. Indeed, if one were to solve the dynamics of the system exactly, one would certainly find that the canonical momenta exhibit altogether different physical characteristics in different gauges, and one would find the amplitudes associated with transitions between bare states to be similarly gauge dependent. A paradigmatic example of this occurrence is found by comparing the field conjugate momentum in the Coulomb and Poincar\'{e} gauges. In the Coulomb gauge ${\bf \Pi}_{\rm T}$ is equal to (the negative of) the non-local transverse electric field $-{\bf E}_{\rm T}$, but in the Poincar\'{e} gauge it is equal to the causally propagating transverse displacement field $-{\bf D}_{\rm T}$.

The experiments of Lamb were in sufficiently close agreement with the Poincar\'{e} gauge result to rule out the Coulomb gauge result. A simple explanation for this is that the physical degrees of freedom represented by the canonical operators in the Poincar\'{e} gauge are closer to the correct ones. This is essentially the explanation first offered by Power and Zienau \cite{PandZ}. The authors of Ref. \cite{milonni} show that if a ``sudden switching" condition of assuming a sharp bare state at $t=0$ is avoided, the Poincar\'{e} gauge lineshape result can be found using the Coulomb gauge. The proposed method to avoid this condition is tantamount to using the canonical observables of the Poincar\'{e} gauge within the Coulomb gauge, so that it essentially resolves the discrepancy between the results in the same way as Ref. \cite{PandZ}.

\section{The atom's excitation by a laser pulse}\label{5}

In this section we choose the Weisskopf-Wigner approach to lineshape derivations \cite{ww}, because it is easily adapted to include a description of the atom's excitation by laser light. Although this treatment does not really allow for any frequency variation of the lineshape numerator in a self-consistent manner, it yields the same results as the resolvent method used in Secs. \ref{2} and \ref{3} provided the Lamb shift is suppressed in the latter. Since our previous lineshapes in Secs. \ref{2} and \ref{3} only depend on one transition frequency $\omega_{eg}$, and we have seen that the quadratic parts of the interaction Hamiltonian do not contribute, we adopt a two-level model for the atom, and a purely linear atom-field interaction Hamiltonian.

\subsection{A simplified version of the Hamiltonian}

Ignoring the laser for the time being, the Hamiltonian we are going to use is found by acting with the transformation in Eq. (\ref{Uv}) on the electric dipole approximated Coulomb gauge Hamiltonian. After neglecting any quadratic interaction terms and restricting ourselves to a two-level atom with transition frequency $\omega_0$, we obtain the interaction Hamiltonian \cite{me2}
\begin{eqnarray}\label{harb}
V&=& \sum_{{\bf k}\lambda } g_{{\bf k} \lambda} \, \sigma^+ \left( u_k^+ \, a_{{\bf k}\lambda }^\dagger + u_k^- \, a_{{\bf k}\lambda } \right) + {\rm H.c.}
\end{eqnarray}
where
\begin{eqnarray} \label{us}
u_k^\pm &\equiv & (1-\alpha_k) \left( {\omega_0 \over \omega_k} \right)^{1/2} \mp \alpha_k \left( {\omega_k \over \omega_0} \right)^{1/2}
\end{eqnarray}
and $\alpha_k$ is a dimensionless function of $\omega_k$ and $\omega_0$. The choices $\alpha_k\equiv 0$ and $\alpha_k=1$ give the Coulomb gauge and Poincar\'e gauge interactions respectively, whereas choosing $\alpha_k$ as in Eq. (\ref{alpha}) gives the symmetric interaction. 

\subsection{Modelling the laser}

To describe the laser we add an appropriate semi-classical interaction term to Eq. (\ref{harb}). For consistency the laser should be taken to couple to the atom in the same way as the quantised field does. Thus we define the atom-laser interaction by
\begin{eqnarray}\label{hlas}
V_l= {i\Omega(t) \over 2} \sigma^+ \left( u_l^+ e^{i\omega_l t} + u_l^- e^{-i\omega_l t} \right) + {\rm H.c.},
\end{eqnarray}
where $\Omega(t)$ is a real but otherwise completely arbitrary time dependent coupling envelope. Assuming resonant driving $\omega_l =\omega_0$, Eq. (\ref{hlas}) reduces to
\begin{eqnarray}\label{hlas2}
V_l= {i\Omega(t) \over 2} \sigma^+ \left( (1-2\alpha) e^{i\omega_0 t} + e^{-i\omega_0 t} \right) + {\rm H.c.},
\end{eqnarray}
where $\alpha$ is an arbitrary real number. Choosing $\alpha = 0$ and defining $\Omega(t) = (e/m) \langle {\bf p}\rangle \cdot {\bf A}_0(t)$, Eq. (\ref{hlas2}) gives a (semi-classical) Coulomb gauge interaction in which the electron's canonical momentum couples to a classical vector potential of the form ${\bf A}_0(t)\cos\omega_0t$;
\begin{eqnarray}\label{hlascg}
V_l= i{e\over m} \langle {\bf p}\rangle \cdot {\bf A}_0(t)(\sigma^+ - \sigma^-)\cos\omega_0t.
\end{eqnarray}
Choosing $\alpha=1$ and defining $\Omega(t) = \langle{\bf d}\rangle \cdot {\bf E}_0(t)$ yields an interaction in which the electron's position (dipole moment) couples to a classical electric field of the form ${\bf E}_0(t)\sin\omega_0t$;
 \begin{eqnarray}\label{hlaspg}
V_l= \langle {\bf d}\rangle \cdot {\bf E}_0(t)(\sigma^+ + \sigma^-)\sin\omega_0t.
\end{eqnarray}
The appropriate choice to accompany the symmetric representation is $\alpha=1/2$.

Before continuing to derive the lineshape we wish to make a note on the rotating wave approximation, which we will use in the following section. The RWA for the atom-laser interaction constitutes the prescription $u^+_l = 0$, which of course holds as an identity in the symmetric representation. If it is to be made in the atom-field interaction use of the RWA ($u_k^+ =0$) may be questionable. This is because in conjunction with the sum over modes the counter-rotating terms give rise to divergent contributions. However, for a single mode or a semi-classical interaction no sum over modes is present and the counter-rotating contributions promise to be very small. When used together with the resonant driving assumption it is clear from Eq. (\ref{hlas2}) that the RWA gives an atom-laser interaction which does not depend on $\alpha$.

\subsection{Calculation of the lineshape}

To derive the lineshape we assume an initial state $\ket{g;0}$. At $t=-\pi/\Omega$ a laser $\pi$-pulse irradiates the atom until $t=0$. If we assume $\Omega \gg \Gamma$ then we can ignore spontaneous emission over the duration of the pulse and set $g_{{\bf k}\lambda} \equiv 0$. For $t\geq0$ the laser has ceased and we make the Weisskopf-Wigner ansatz of exponential decay of the excited atomic state; $b_{e;0}=e^{-\Gamma t/2}$. With these assumptions in place we calculate the long time amplitude $b_{g;{\bf k}\lambda}(\infty)$ as in Sec. \ref{2}. 

\begin{center}
\begin{figure*}[t]\label{fig2} 
\subfigure[\hspace*{-1.8mm}]
{\includegraphics[width=0.48\textwidth]{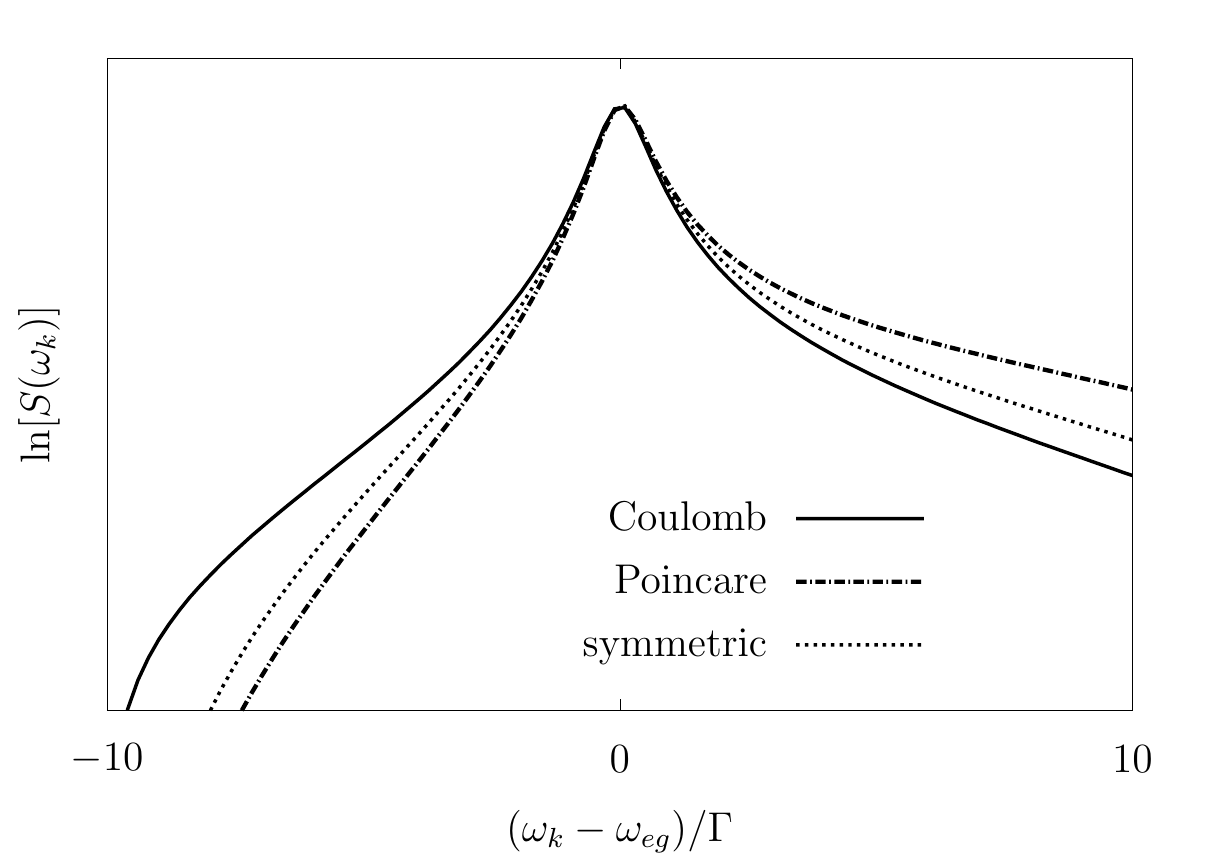}\label{laser1}}
\subfigure[\hspace*{-2.7mm}]
{\includegraphics[width=0.48\textwidth]{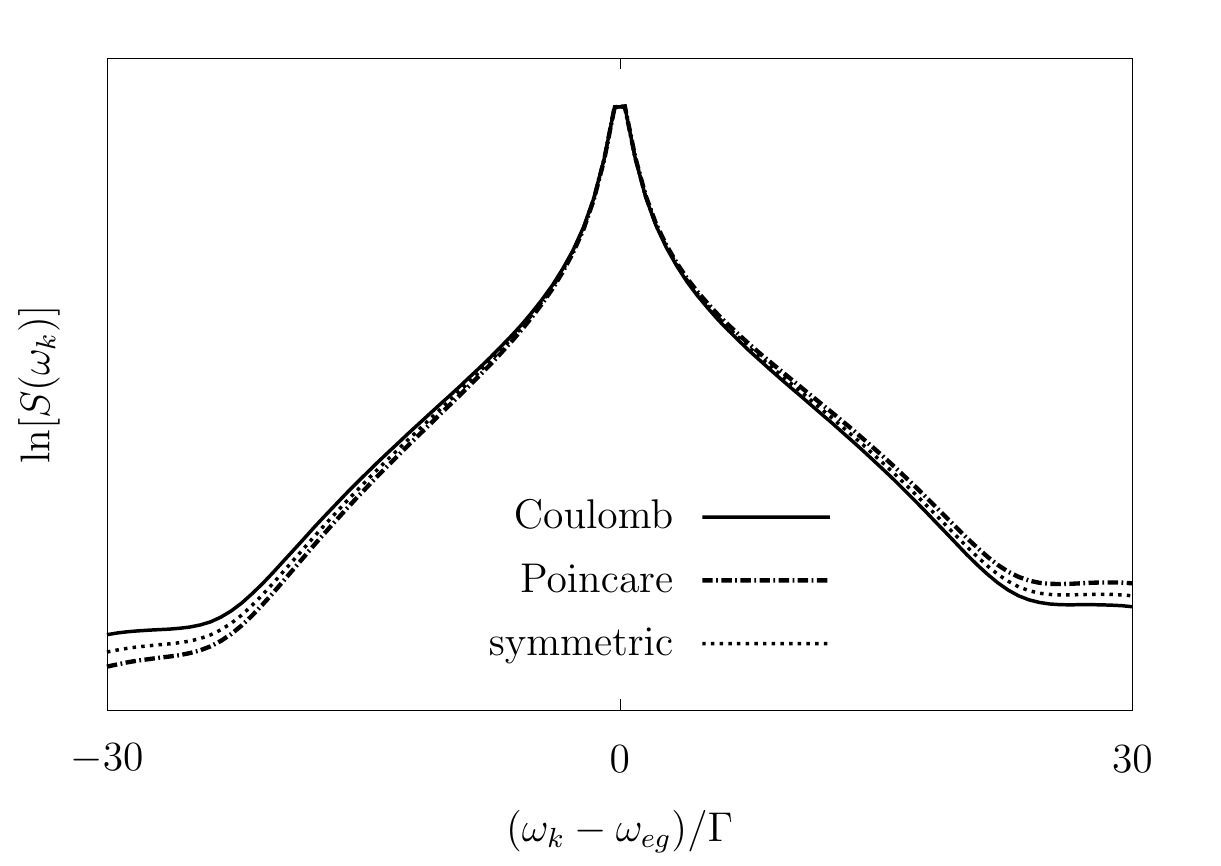}\label{laser2}}
\subfigure[\hspace*{-1.8mm}]
{\includegraphics[width=0.48\textwidth]{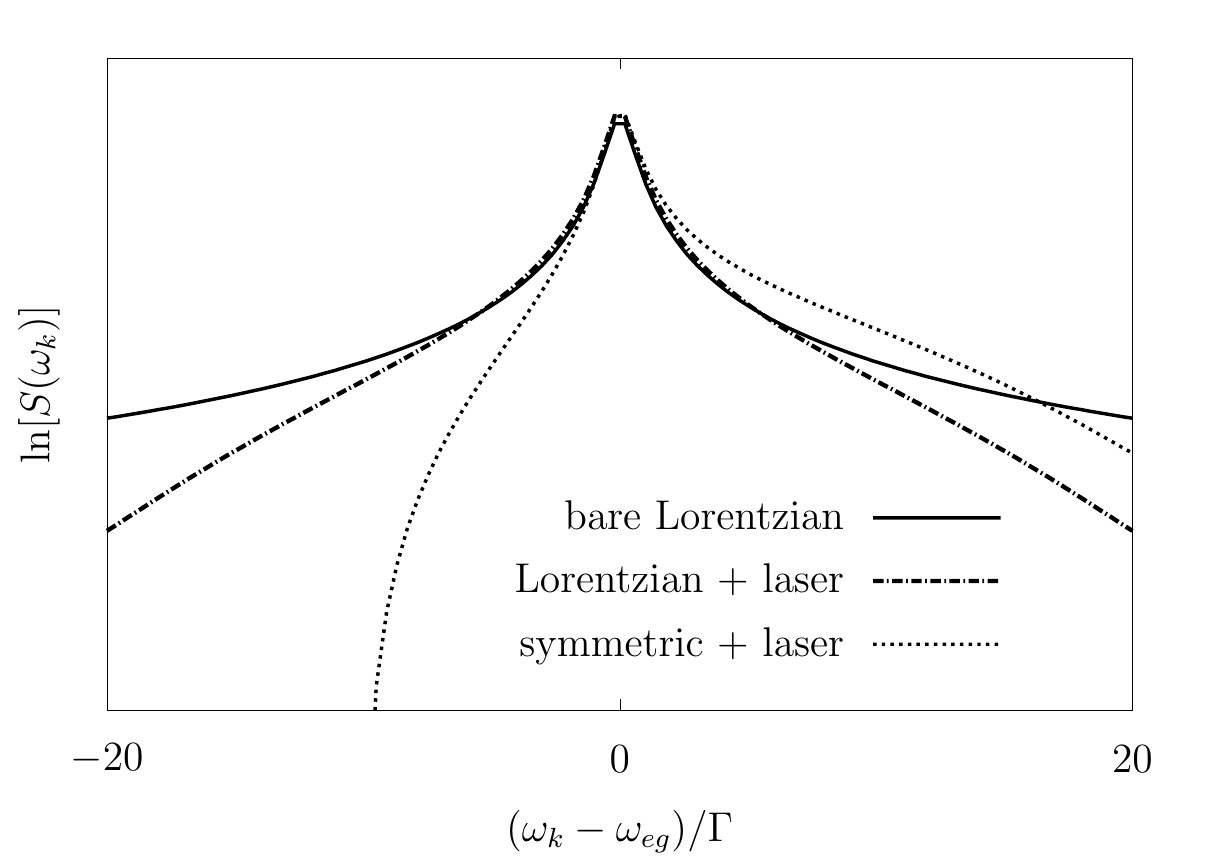}\label{laser3}}
\hspace*{0.5cm}\subfigure[\hspace*{-2.7mm}]
{\includegraphics[width=0.48\textwidth]{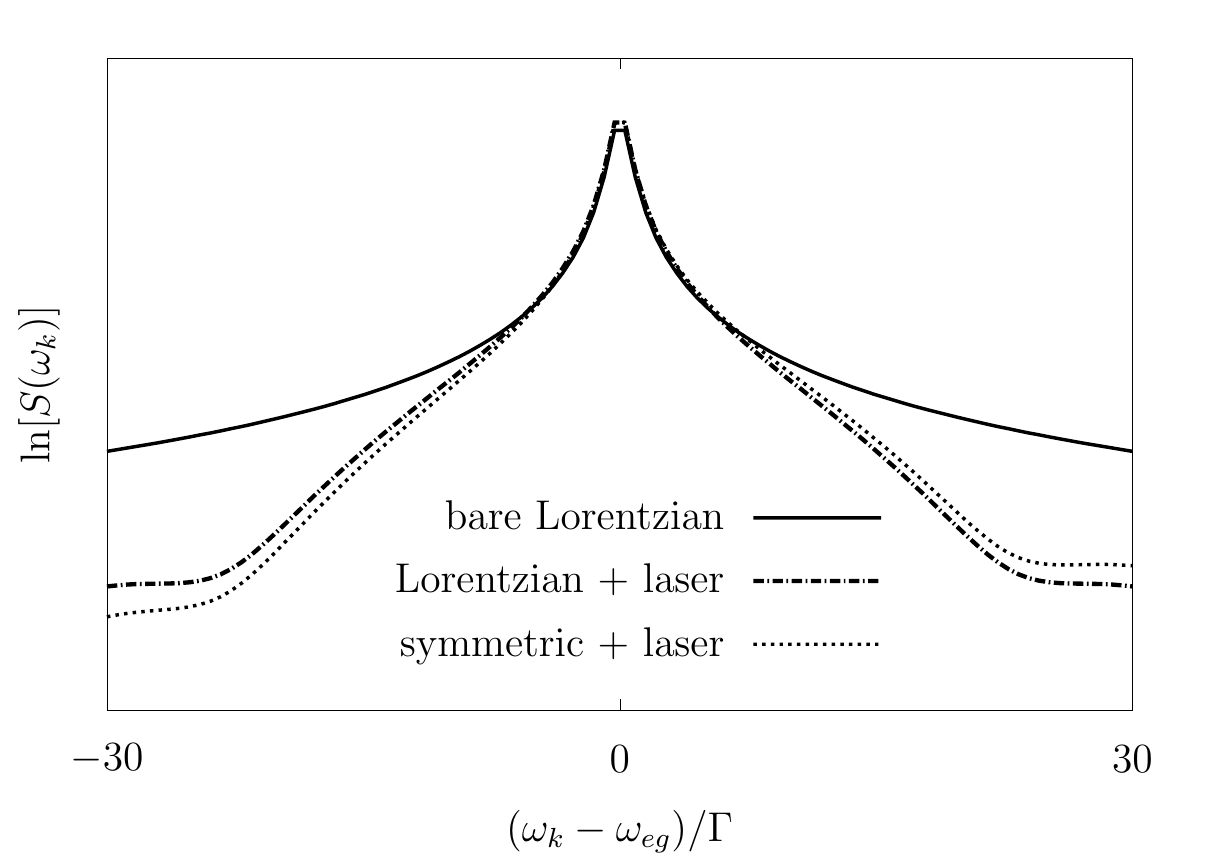}\label{laser4}}
\vspace*{-0.2cm}
\caption{\small{$\Omega=\omega_0$ and $\delta_l=0$. In (a) and (b) the lineshapes associated with the Coulomb gauge, Poincar\'e  gauge and the symmetric representation are plotted, each lineshape includes the laser contribution. In (a) $\Gamma = \omega_0/10$ and in (b) $\Gamma = \omega_0/100$. In (c) and (d) the lineshape in the symmetric representation including the laser contribution is compared to the bare Lorentzian curve  $(\Gamma /2\pi)/(\delta_k^2+\Gamma^2/4)$, and to the Lorentzian including the laser contribution.  In (c) $\Gamma = \omega_0/10$ and in (d) $\Gamma = \omega_0/100$.}}
\end{figure*}
\end{center}

We will consider the rectangular envelope
\begin{eqnarray}
\Omega(t) = \left\{ \begin{array}{rl}
 \Omega &\mbox{ if $-\pi/\Omega <t<0$} \\
  0 &\mbox{ otherwise},
       \end{array} \right.
\end{eqnarray}
and begin with the assumption that the state at time $t$ can be expanded as
\begin{eqnarray}\label{psi}
\ket{\psi (t) } = b_{g;0}\ket{g;0}e^{-i\omega_g t} + b_{e;0}\ket{e;0}e^{-i\omega_e t} + \sum_{{\bf k}\lambda} b_{g;{\bf k}\lambda}\ket{g;{\bf k}\lambda}e^{-i(\omega_g + \omega_k)t}.
\end{eqnarray}
Adding the atom-field interaction in Eq. (\ref{harb}) to the atom-laser interaction in Eq. (\ref{hlas}) and using Eq. (\ref{psi}), the Schr\"odinger equation yields the set of coupled differential equations
\begin{eqnarray}\label{diff}
{\dot b}_{g;{{\bf k}\lambda}} &=& -ig_{{\bf k}\lambda}^*u_k^-e^{-i(\omega_0-\omega_k)t}b_{e;0}, \nonumber \\
{\dot b}_{g;0} &=& -{\Omega(t)\over 2}(u_l^+ e^{-i\omega_l t}+u_l^- e^{i\omega_l t})e^{-i\omega_0 t}b_{e;0}, \nonumber \\
{\dot b}_{e;0} &=& {\Omega(t)\over 2}(u_l^+ e^{i\omega_l t}+u_l^- e^{-i\omega_l t})e^{i\omega_0 t}b_{g;0} -i\sum_{{\bf k}\lambda} g_{{\bf k}\lambda} u_k^- b_{g;{{\bf k}\lambda}}e^{i(\omega_0-\omega_k)t}.
\end{eqnarray}
Setting $g_{{\bf k}\lambda}=0$ for $-\pi/\Omega<t<0$ and implementing the RWA we obtain
\begin{eqnarray}\label{diff2}
{\dot b}_{g;0} &=& -{\Omega(t)\over 2}u_l^- e^{-i(\omega_0-\omega_l) t}b_{e;0}, \nonumber \\
{\dot b}_{e;0} &=& {\Omega(t)\over 2}u_l^- e^{i(\omega_0-\omega_l)t}b_{g;0}.
\end{eqnarray}
With initial conditions $b_{g;0}(-\pi/\Omega) = 1, \, b_{e,0}(-\pi/\Omega) = 0$, Eq. (\ref{diff2}) yields the solution
\begin{eqnarray}\label{soln}
b_{e;0} = -{i\Omega u_l^-\over \mu}e^{i\delta_l (t-\pi / \Omega)/2}\sin\left[{\mu \over 2}\left(t+{\pi\over \Omega}\right)\right],
\end{eqnarray}
where $-\pi / \Omega< t < 0$ and
\begin{eqnarray}\label{mudel}
\delta_l \equiv  \omega_0-\omega_l, ~~~~~
\mu \equiv  \sqrt{(\Omega u_l^-)^2 + \delta_l^2}.
\end{eqnarray}
Substituting this solution into Eq. (\ref{diff}), along with $b_{e;0}=e^{-\Gamma t/2}$ for $t\geq0$, and then integrating with respect to $t$ yields the result
\begin{eqnarray}\label{b}
&& \hspace*{-0.6cm} b_{g;{\bf k}\lambda} (\infty) = -ig_{{\bf k}\lambda}^*u_k^-  \nonumber \\ && \hspace*{-0.6cm}  \cdot \bigg\{ {1 \over i\delta_k + \Gamma /2} + {2u_l^-\Omega e^{-i\pi\delta_l / 2\Omega} \over (\Omega u_l^-)^2 + 4\delta_k \delta_{kl}}\bigg[ e^{i\pi (2\delta_{k}-\delta_l) / 2\Omega}-\cos\left({\pi\mu \over 2\Omega}\right) - {i\over \mu}(2\delta_k-\delta_l)\sin\left({\pi\mu\over 2\Omega}\right)\bigg]\bigg\}, \nonumber \\
\end{eqnarray}
where $\delta_k \equiv \omega_0-\omega_k$ and $\delta_{kl} \equiv \omega_k-\omega_l$. Unless the laser driving is resonant $b_{g;{\bf k}\lambda} (\infty)$ depends on the laser detuning, and therefore on the form of the atom-laser coupling through $u_l^-$. It happens that this dependence is actually extremely weak, so the only notable dependence on the laser comes through $\Omega$. A great simplification is afforded by assuming a resonant pulse whereby Eq. (\ref{b}) reduces to
\begin{eqnarray}\label{b2}
b_{g;{\bf k}\lambda} (\infty)  = -ig_{{\bf k}\lambda}^*u_k^-\bigg[ {1 \over i\delta_k + \Gamma /2 } + { 2 \over \Omega^2 - 4\delta_k^2} \bigg( \Omega e^{i\pi \delta_k / \Omega} -2i \delta_k  \bigg)\bigg].
\end{eqnarray}
The lineshape is defined in Eq. (\ref{S}). Near resonance the Lorentzian component dominates, but in the wings the lineshape is sensitive to $\Omega$. As in Sec. \ref{2} it is dependent on the representation chosen through the function $u_k^-$. Various lineshapes including the laser contribution are plotted in Figs. \ref{laser1}-\ref{laser4}. For an optical transition $\omega_0 \sim 10^{-15}$ with decay rate $\Gamma \sim 10^8$ the differences in lineshapes associated with different representations are extremely small, significant differences only occurring for much larger decay rates relative to $\omega_0$. This situation may be improved with the use of different laser pulse envelopes. We note that even for these parameters, the difference between lineshapes with and without taking into account the laser should be detectable with modern spectroscopy.

%
%
%
%

 \section{Conclusions}\label{6}

The gauge dependence of lineshapes found using conventional approaches has been investigated. Particular representations of the atom-field Hamiltonian, which may be of special physical significance have been considered in detail. As well as discussing the necessary and sufficient conditions that the lineshape be gauge invariant we have determined the specific shapes for the cases of interest. In order to put these results within the context of realistic experimental conditions, we have extended our investigation to include an account of the atom's excitation. We have attempted to include a reasonably broad range of calculational techniques by using both the formal theory of radiation damping as well as the traditional Weisskopf-Wigner treatment. An experiment might reveal which of the calculated lineshapes is closest to the measured one.

{\em Acknowlegdement}. The author would like to thank A. Beige and T. P. Spiller for useful discussions regarding this work.

\end{document}